\documentclass[aps,prl,twocolumn,showpacs,preprintnumbers,amsmath,amssymb]{revtex4-2}
\usepackage{graphicx}% Include figure files
\usepackage{dcolumn}% Align table columns on decimal point
\usepackage{bm}% bold math
\begin{document}
\title{BU~CMi as a quadruple doubly eclipsing system}%

%\correspondingauthor{Igor M. Volkov} \email{hwp@yandex.ru}

\author{Igor M. Volkov}
\affiliation{Sternberg Astronomical Institute, \\
Lomonosov Moscow State University, \\
Universitetskii pr.13, 119991 Moscow, Russia}

\affiliation{Institute of Astronomy of the Russian Academy of Sciences, \\
Pyatnitskaya str.48, 119017 Moscow, Russia}

\author{Alexandra S. Kravtsova}
\affiliation{Sternberg Astronomical Institute, \\
Lomonosov Moscow State University, \\
Universitetskii pr.13, 119991 Moscow, Russia}

\affiliation{Institute of Astronomy of the Russian Academy of Sciences, \\
Pyatnitskaya str.48, 119017 Moscow, Russia}

\author{Drahomir Chochol}
\affiliation{Astronomical Institute of the Slovak Academy of
Sciences, 059 60 Tatransk\'a Lomnica, Slovakia\\
              \email{chochol@ta3.sk}
             }

\date{\today}

\begin{abstract}
 We found that the known spectroscopic binary and
variable BU~CMi = HD65241 ($V$=6.4-6.7 mag, Sp~=~A0~V) is a
quadruple doubly eclipsing 2+2 system. Both eclipsing binaries are
detached systems moving in an eccentric orbits: pair "A" with the
period $P_A$~=~$2^{d}.94$($e$=0.20) and pair "B" with the period
$P_B$~=~$3^{d}.26$ ($e$=0.22). All four components have nearly
equal sizes, temperatures and masses in the range $M$~=~3.1--3.4
M$_\odot$ and $A0$ spectra. We derived the mutual orbit of both
pairs around the system barycenter with a period of 6.54 years and
eccentricity $e$ = 0.71. We detected in pairs "A" and "B" the fast
apsidal motion with the periods $U_A$~=~25.0 years and
$U_B$~=~25.2 years, respectively. The orbit of each pair shows
small nutation-like oscillations in periastron longitude. The age
of the system estimated as 200 mln. years. The photometric
parallax calculated from the found parameters coincides perfectly
with the $GAIA~DR2$ $\pi$=$0.00407"\pm0.00006"$.
\end{abstract}


\maketitle

\section{Introduction}

\begin{figure}
\includegraphics[width=\columnwidth]{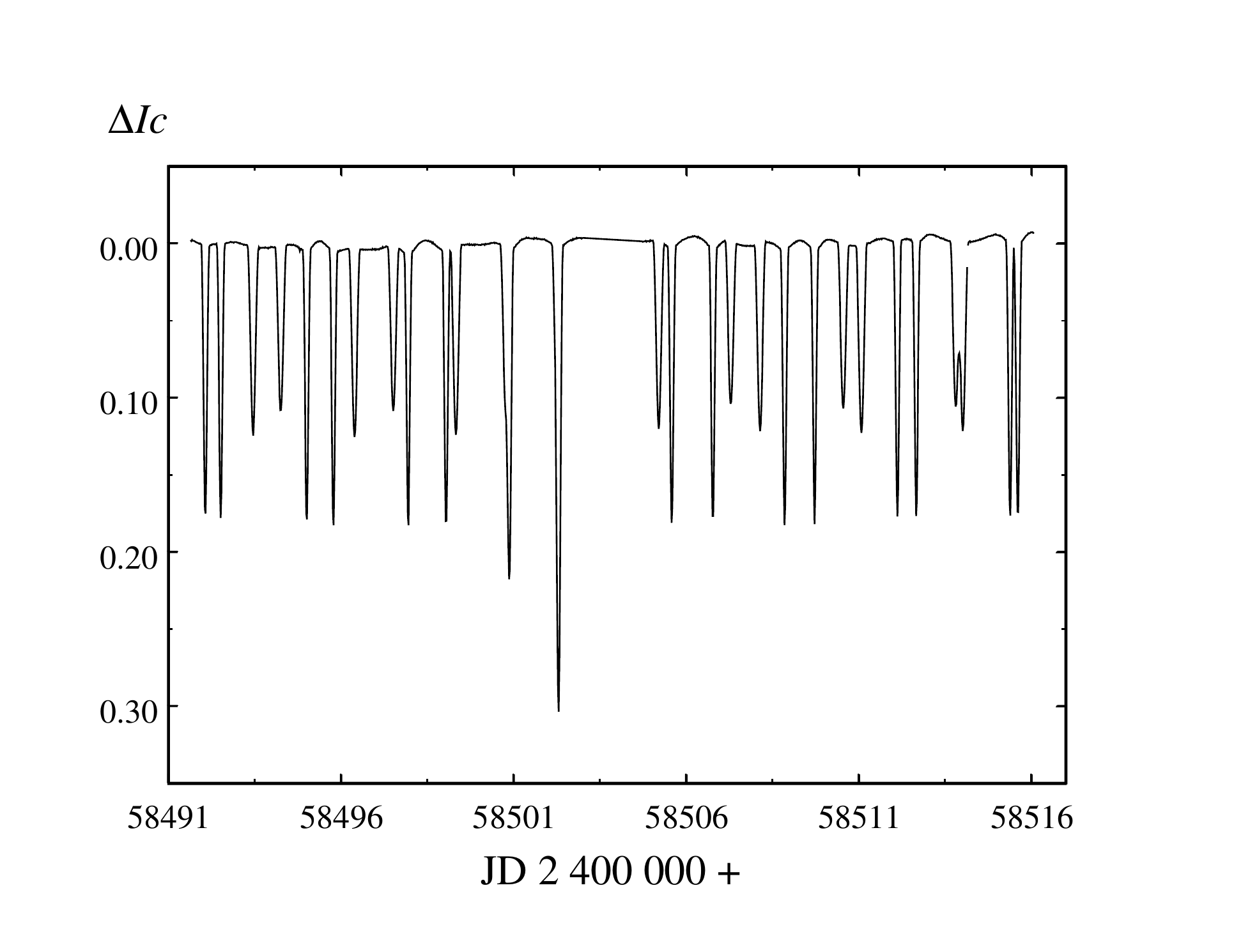}
\caption{TESS observations of BU~CMi. Two groups of eclipses of
stars "A" and "B" with stable periods and displaced secondary
minima due to eccentric orbits are detected.
              }
\label{Tess}
\end{figure}

The system BU~CMi is designated in GCVS as an eclipsing variable
EA with the period P~=~$2.93^d$ and strongly displaced secondary
minimum MinII - MinI = 0.37P, \citet{2017ARep...61...80S}, which
means the large eccentricity of the orbit. So, it was included in
our program of investigation of the inner structure of the stars
\citet{2009ARep...53..136V}. We started the observations in 2012
at Star\'a Lesn\'a observatory, Slovakia. We managed to fix only
one exit from minimum strongly apart of the GCVS ephemeris. Our
further observations in the same year did not bring any result,
the depth of the only observed eclipse turned out to be
insignificant and the system was classified as unpromising for the
inner structure investigation. But when the reviews of the bright
stars such as MASCARA \citet{2018A&A...617A..32B} and TESS
\citet{2019AJ....158..138S} became available we returned to
analysis of the system. The ASAS data \citet{2002AcA....52..397P}
for this bright star were useless as their precision was poor, may
be due to overexposure. The first glance to the Light Curve (LC)
built with MASCARA data revealed that there are two groups of
eclipses with different but stable periods and displaced secondary
minima (see Fig.~\ref{Tess} with more illustrative TESS
observations). Using the corrected ephemeris obtained from MASCARA
observations we have resumed intensive observations of the star at
the Simeiz observatory of INASAN with the 60-cm reflector and
$UBV$-photometer constructed by I.M.Volkov,
\citet{2007A&AT...26..129V}. Here we present the corrected results
firstly published in \citet{2021ARep...98..115V}. We managed to
correct those results by disentangling far removed from the
current epoch Hipparcos data, \citet{1997A&A...323L..49P}.
%
%-------------------Figure 1 --------------------------
\section{Observations and data reduction}

The log of our observations is presented in Table~\ref{journal}.
During observations in 2020 all four minima - two primary and two
secondary ones, were recorded for both eclipsing systems observed
simultaneously as a single star. HD64963 ($V$=8.23, A2) at a
distance 22' of the variable served as a standard star for
observations with $UBV$-photometer equipped by photomultiplier.
The observations were carried out according to the standard star -
variable star - standard star scheme. The expositions in each of
the photometric passbands were from 20 to 30 sec, sometimes the
background near the standard and variable stars was measured. We
used aperture 27.5". The maximum value of the signal from the star
did not exceed 200000sec$^{-1}$ in $B$. The signal was weaker in
other photometric passbands. Such rate of the signal needs a
correction due to the nonlinearity of the equipment not more than
$0^m.005$, which was taken into account according to the
well-known nonlinearity formula for the Poisson distribution of
impulses in the photon-counting flux. The dead time of our
registration channel: photomultiplier + amplifier + pulse counter
averaged 32.7 nanoseconds for the entire period of observations.
This value has been carefully controlled, it remained stable
within 5 percent. All observations were corrected for atmospheric
extinction. More details of our observational method are described
in \citet{2021ARep...65..184V}, the instrumental system is
presented in \citet{2007A&AT...26..129V}. Since the number of
nights suitable for observations with a single-channel photometer
is limited by weather conditions, we attempted to use CCD
observations with standard and variable stars measured
simultaneously. To attenuate the peak signal from such a bright
star as BU~CMi, images were defocused and exposures were
shortened. The result was unsatisfactory. We failed to achieve the
required accuracy better than $0^m.01$ and the shape of the minima
was distorted by systematic errors.  The CCD observations in
Star\'a Lesn\'a observatory, Slovakia were much better. We used a
Maksutov with a 15-cm aperture, so it was not necessary to
attenuate the signal artificially. In addition, a larger field of
view ensured the presence of a bright comparison star in one frame
with a variable star. In the spring 2021 we obtained at the Simeiz
observatory of INASAN only the incomplete observations in four
minima due to unsatisfactory weather conditions. But these
observations turned out to be very important for determination of
the parameters of the external orbit of A+B pairs.

BU~CMi was also measured six times during observations of the SAI
catalogue of bright stars \citet{1991TrSht..63....4K}. We
recalculated the ultraviolet $W$ values given in this catalogue
into the standard $U$ system, and the instrumental values $B$, $V$
were corrected a little to correspond precisely to Johnson system.
Two of these measurements were obtained in minima, which enable us
to fix the moment of the minimum farthest from the current epoch,
and four measurements turned out to be located between the minima
and were averaged. Data of the brightness of BU~CMi on the plateau
and comparison star are given in Table~\ref{magnitudes}.

When analyzing the observations of Hipparcos
\citet{1997A&A...323L..49P}, we noticed that some of the
observations of the object fall into minima, but, as a rule, not
more than one or two. These observations are slightly less distant
from the current epoch than the SAI catalogue observations and we
were not able to use them prior to building the model of the
system published in \citet{2021ARep...98..115V}. Now we are in a
position to correctly classify each of the brightness attenuation
and supplement our analysis with these data. We found that the
observed brightness attenuations correspond to four secondary and
one main minima of the "B" component.

Also, at our request, T.~Pribulla and R.~Kom\v{z}ik carried out
spectroscopic observations of the system using the high-resolution
echelle spectrograph mounted on 1.35-m reflector at the Skalnat\'e
Pleso observatory, Slovakia \citet{2019CoSka..49..154D}. Obtained
spectra contain moving lines of four different stars, with the
spectral types close to A0 and approximately of the same
brightness. We used the spectra to construct the radial-velocity
curves (RVCs) for all four components of the system BU~CMi.

\section{Interstellar extinction and temperatures of the components}

The star under investigation is at a distance of 250 parsecs from
us and is located far enough from the Galactic equator,
$b~=~18^\circ$, so we cannot expect large interstellar absorption.
This is confirmed by the position of the star on the two-colour
$U-B,B-V$ diagram. We do not show it here, but a similar one can
be seen in Fig.~3 of our previous work
\citet{2021ARep...65..184V}. The colour indices of the standard
star HD64963 do not show any significant interstellar reddening,
too. According to \citet{2015ApJ...810...25G}, zero interstellar
absorption in the direction of BU~CMi follows up to a distance of
several kiloparsecs. The maps of interstellar reddening
\citet{1998ApJ...500..525S}, \citet{2011ApJ...737..103S} suggest
for BU~CMi interstellar extinction $E(B-V)$=$0^m.007\pm0.0006$.
Since it is difficult to determine the interstellar extinction
with the same accuracy from our two-colour $U-B,B-V$ diagram, we
accepted for further analysis this value of extinction and
corrected the star colour index from Table~\ref{magnitudes}.
Despite the fact that reddening is close to zero, its value in the
range of the colour indices corresponds to a temperature
difference of 140K according to calibration used
\citet{1980ARA&A..18..115P}, \citet{1996ApJ...469..355F}.

It is interesting to note that although some young elliptical
systems, such as GG~Ori \citep{2002ARep...46..747V}, V944~Cep
\citep{2015ASPC..496..266V}, V2544~Cyg
(\url{https://www.eso.org/sci/meetings/2017/ImBaSE2017/Posters/Poster_17-Chochol.pdf})
and V839~Cep \citep{2019CoSka..49..434V} exhibit anomalously large
absorption, BU~CMi does not show any significant deviations in
absorption.

\section{The photometric analysis of the system}

In our studies, we solve the complex problem of determining the
entire set of interconnected characteristics of a multiple system,
and the problem is complicated by the fact that there are two
eclipsing binaries and the brightness of all four stars is
measured simultaneously. Thus, the number of free parameters is at
least double. A number of natural conditions limit the range of
possible parameters of the model, which facilitates its
construction. These conditions are as follows: First, the combined
flux of the system must match the colour indices given in
Table~\ref{magnitudes}. Second, Kepler's laws must be fulfilled
for each of the two eclipsing pairs separately, and for the
external orbit for both pairs together. There may be some
deviations from these laws, because each of the systems cannot be
considered closed, the stars themselves cannot be considered as
material points, as evidenced by the apsidal motion and small
brightness variations between minima, see Fig.~\ref{Tess}. But as
a first approximation, the laws should work. Third, the
photometrically determined distance to all four stars should be
the same. Fourth, the total brightness of all four stars is taken
as unity. The last condition immediately implies that we will
solve the LCs with a significant contribution of the third light.
Since the depths of all four minima are approximately the same and
close to $0^m.2$, and the power of all lines is also approximately
the same, it can be expected that all four stars have a similar
surface brightness or, equivalently, temperature. We also assume
that the empirical mass-luminosity law is fulfilled.

The applied technique is described in detail in a number of our
previous articles, such as \citet{2021ARep...65..184V},
\citet{2017ARep...61..440V}, \citet{2018ARep...62..664B},
\citet{2019ARep...63..495K} and \citet{2020ARep...64..211V}. Here,
we just note that the stellar disk is represented by an ideal
circle, the brightness distribution over it is taken into account
by concentric circles of different brightness,
\citet{1984SvA....28..228K}, in the model of the linear law of
darkening to the edge, and the final solution is obtained by the
differential corrections method. Edge-darkening coefficients were
fixed from the theoretical models \citet{1985A&AS...60..471W}.
Within the framework of this model, there are no brightness
variations between the minima. The analysis revealed that in very
accurate TESS observations, see Fig.~\ref{Tess}, such changes at
the level of fractions of a percent exist, but within the errors
of the determined parameters, they do not affect the solutions at
the minima.

\begin{figure}
\centering
\includegraphics[width=\hsize]{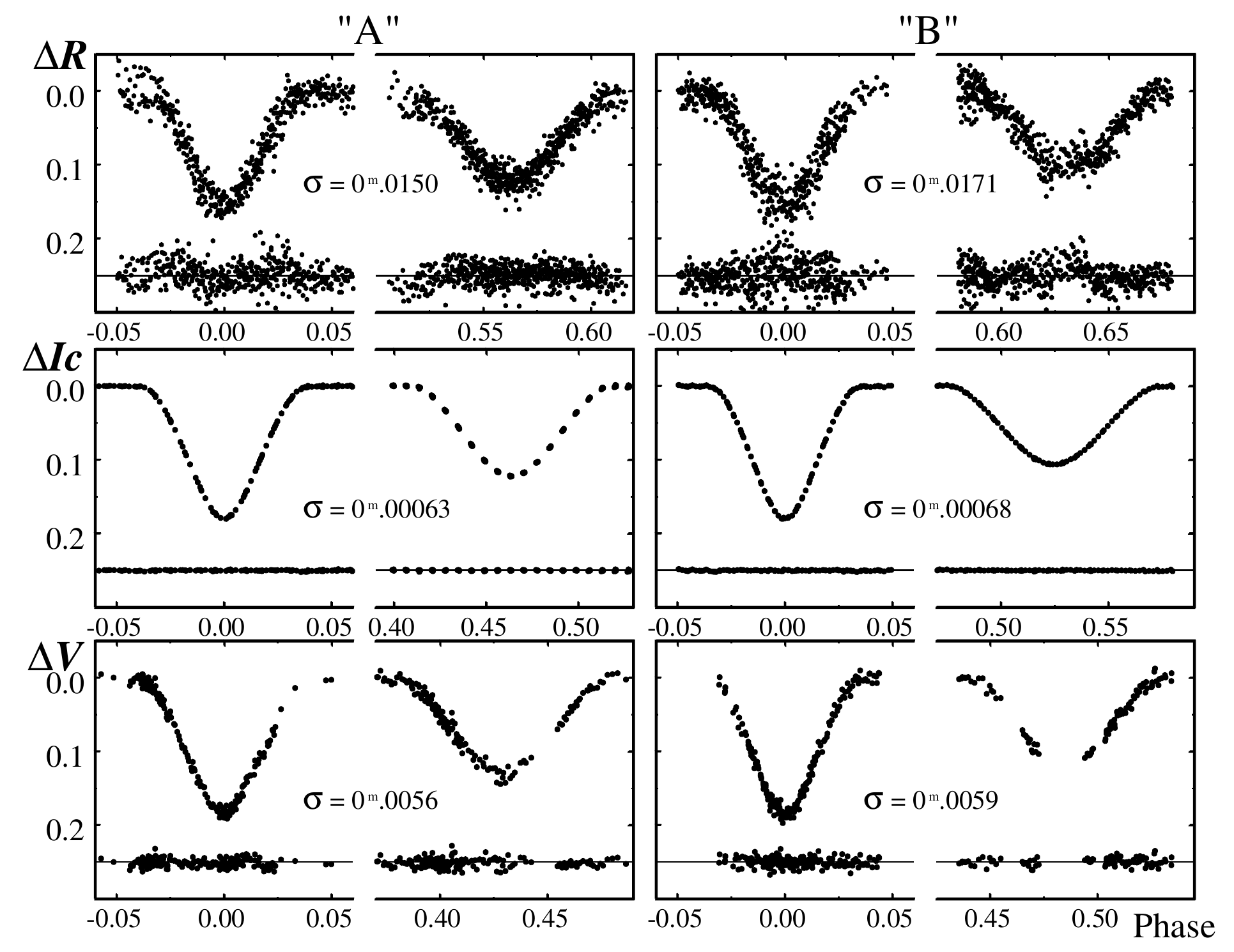}
\caption{Observations in minima. Left panels: component "A", right
panels: component "B". The plateau level is assigned to zero. The
$O~-~C$ residuals from   concrete solution are presented under
each minimum. The average error value is indicated. Observations
are located in chronological order from top to bottom. Top row:
MASCARA observations, middle row: TESS observations, lower row:
our observations in the $V$ passband. A progressive decrease of
the phases in the secondary minima  caused by the fast apsidal
movement is clearly seen.
              }
\label{solutions}
\end{figure}

There are three independent photometric sets in our disposal. The
first set is the MASCARA unfiltered observations
\citet{2018A&A...617A..32B}. We solved them under the assumption
that the average wavelength of the observations corresponds to the
Johnson $R$ band. The second set comprise our own $UBV$
observations and the third set the TESS observations
\citet{2019AJ....158..138S}, to which the Cousins $Ic$ band was
attributed. Each of the available observational sets was analyzed
separately and the results were compared. We got approximately the
same independent parameters, and the light of "A" and "B"
components was equally divided in all solutions. This is an
important point. At the beginning, we received the same errors for
our $V$ observations and TESS points, at the $0^m.006$ level. But
after a closer look at the residual deviations in TESS
observations, it was found that by adding the small corrections to
the time of observation at each date, the accuracy of the LC
solution can be improved tenfold, from $0^m.006$ to $0^m.0006$.
Later, having plotted on a large scale the graphs of the course of
the $O-C$ residuals at the minima timings, we explained the
introduction of these corrections by nutation, see the section VI.
The real weight of each individual observation increased by a
factor of 100. After correcting the TESS observations for this
effect, we obtained the most accurate solution, which was taken as
the final one. The approximation of our observations and MASCARA
observations by this model does not worsen the scatter of points
on the corresponding LCs. The results for all LCs are shown in
Fig.~\ref{solutions}. All these curves are approximated by the
same parameters, except for the longitude of the periastron, which
changes with time due to the rapid apsidal rotation. The
parameters obtained from the solutions are presented in
Table~\ref{masses}. The masses of the stars were estimated from
photometric data using an indirect method, see
\citet{2017CoSka..47...29V}. The way in which we estimate the mass
errors of the indirect method is described in
\citet{2021ARep...65..184V}. The parameters found from the
photometry served as the basis to disentangle the spectroscopic
observations. Table~\ref{masses} lists the masses already refined
from the RVCs.

The obtained relative luminosities and temperatures, in accordance
with the calibrations \citet{1996ApJ...469..355F}, correspond to
the non-reddened colour index $B-V$ = $-0^m.031$. This value is
larger by $0^m.004$ than that measured in the catalogue
\citet{1991TrSht..63....1K}, see Table~\ref{magnitudes}. We
believe that such a discrepancy is an objective assessment of both
the accuracy of the calibrations \citet{1996ApJ...469..355F} and
the uncertainty of our solution, i.e., the match is satisfactory.

\section{The spectral observations and their analysis}

\begin{figure}
\centering
\includegraphics[width=\hsize]{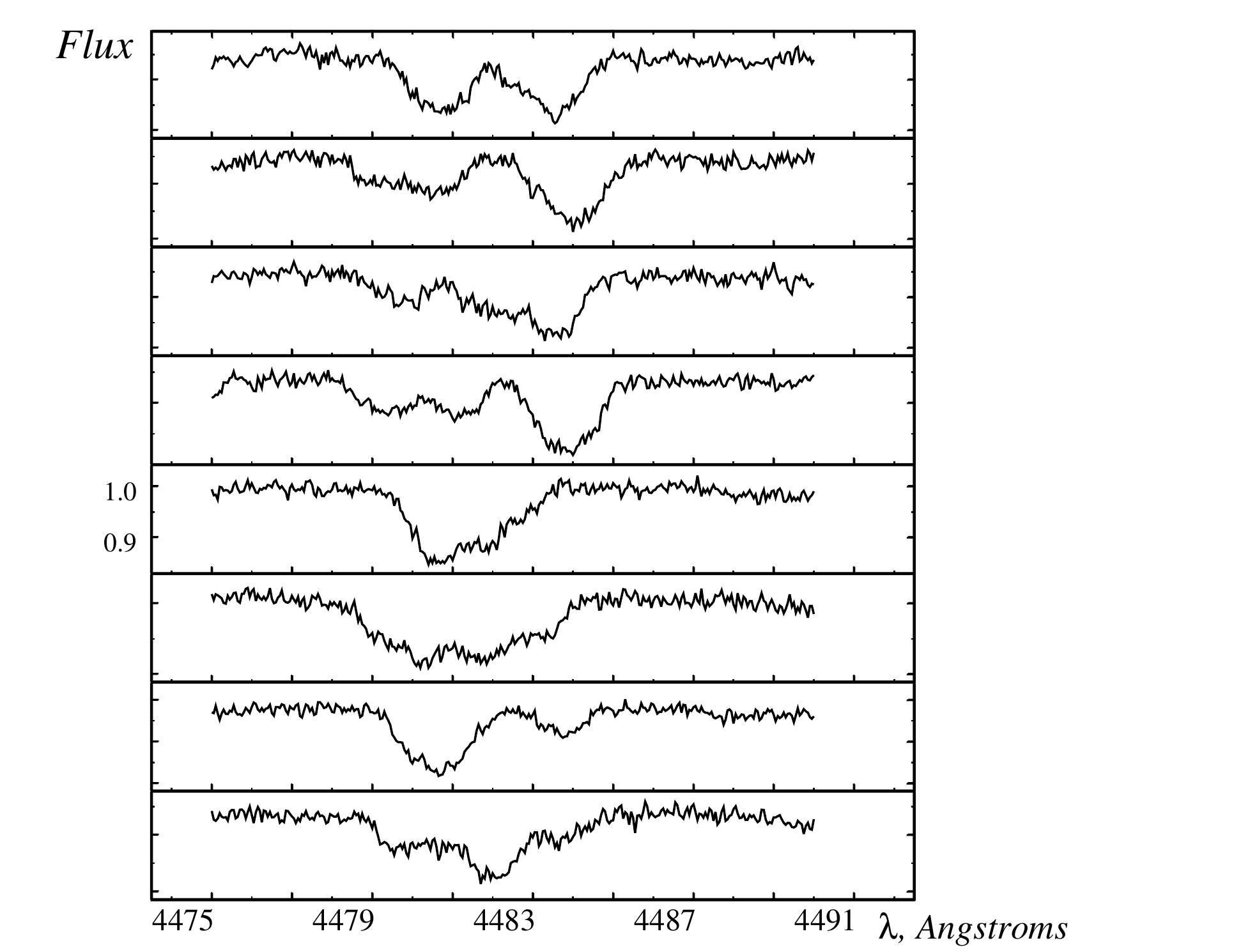}
\caption{The part of BU~CMi spectrum near MgII (4481\AA) line. The
relative flux normalized to the continuum is plotted along the
ordinate axis. The scale shown in one chart is the same for all
charts. Every blend consists of at least two lines, so one can see
the presence of moving lines from four components. These four
lines have approximately the same intensity.
              }
\label{spectra}
\end{figure}

\begin{figure}
\centering
\includegraphics[width=\hsize]{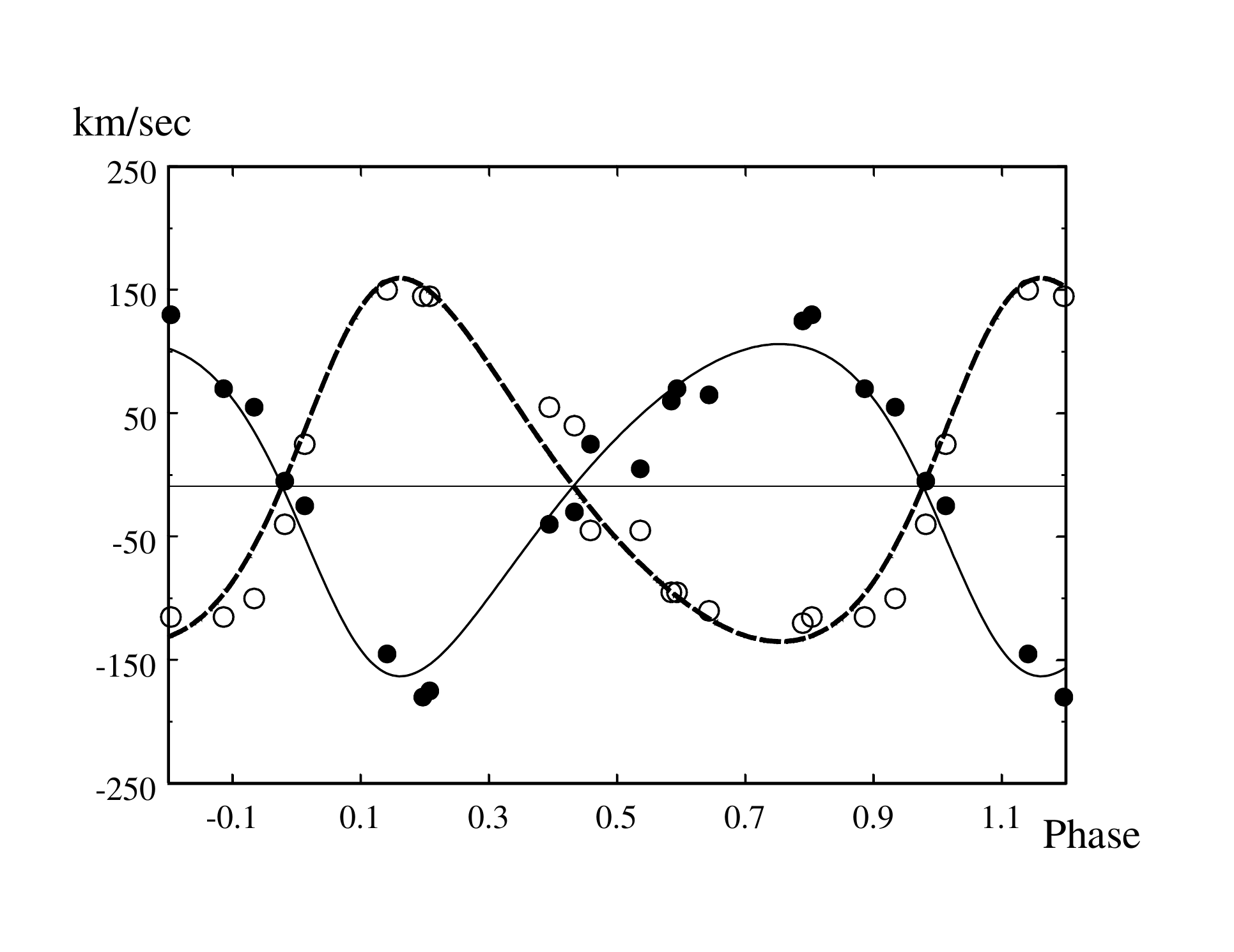}
\caption{Radial velocities of "A" star. Filled and open circles
present the primary and the secondary component, respectively. The
solid line corresponds to the solution of the radial velocity
curve for the primary component and dashed line for the secondary
one. The average error of a separate measurement is 13.3 km/sec.
              }
\label{RV_A}
\end{figure}

\begin{figure}
\centering
\includegraphics[width=\hsize]{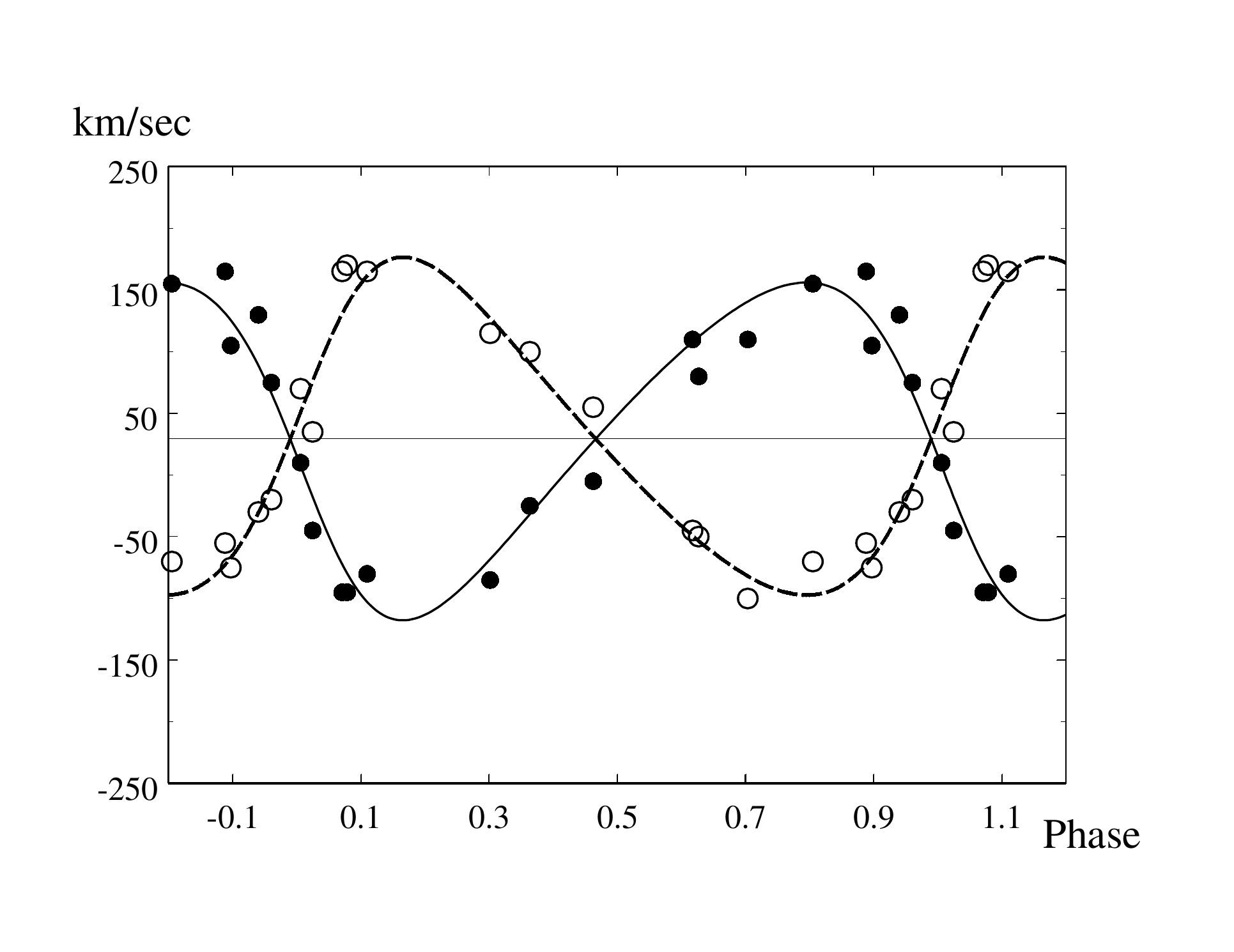}
\caption{Radial velocities of "B" star. Filled and open circles
present the primary and the secondary component, respectively. The
solid line corresponds to the solution of the radial velocity
curve for the primary component and dashed line for the secondary
one. The average error of a separate measurement is 15.5 km/sec.
              }
\label{RV_B}
\end{figure}

Disentangling the spectra in order to measure the radial
velocities of the components turned out to be difficult task,
because the line profile of each component was almost always in
blend with the line profiles of other components.  The line
profiles of all four components were approximated by Gaussians of
the same amplitude and width, and the best fit was sought between
the resulting general approximation function and the observed
spectrum. We used for spectral type A0 the most powerful optically
thin lines MgII (4481\AA), see Fig.~\ref{spectra}. The results of
the analysis are presented in Table~\ref{radial} and in
Fig.~\ref{RV_A},~\ref{RV_B}. The significant scatter is due to the
constant superposition of lines in the spectra. When solving the
radial-velocity curves, the apsidal rotation was taken into
account. The masses of the components are shown in
Table~\ref{masses}. These results are not final; work on the
spectra is ongoing.
\section{The apsidal rotation and the light time effect in minima timings}

To study both effects, it is necessary to know the exact times of
the minima. We have determined all possible timings from the
MASCARA, TESS and our observations. According to the conditions of
the star's visibility above the horizon at the Simeiz observatory
of INASAN, we were unable to observe any full eclipse. From the
observed individual branches of the minima the average minima were
constructed for our observations in 2020. Thus, we have obtained
four average eclipse timings, one for each of the stars in the
system. Another individual moment of minimum was obtained from our
observations at Star\'a Lesn\'a in 2012. The bibliographic search
revealed only one published minimum timing
\citet{2008IBVS.5843....1O}. One more timing we derived from the
observations of the SAI catalogue.
Although there are only two observational points at this minimum,
the amplitude of the brightness attenuation indicates that they
occurred at the time instant corresponding to the bottom of the
minimum. Due to the fact that the observations were obtained in
the most distant time from the modern era it most accurately
determines the period of revolution in the outer orbit. And
finally, from observations of the parts of the minima in 2021
using the established geometric model of the star, the four most
recent minima timings were obtained. For each of the available
timings it was determined which component of the quadruple system
is eclipsed, the blending minima were discarded. All obtained data
are presented in Tables~\ref{timings_A_pri}, \ref{timings_A_sec},
\ref{timings_B_pri}, \ref{timings_B_sec}.

The largest contribution to the deviation of the course of the
minima timings from the linear formula is made by the apsidal
rotation. The amplitude of the effect reaches 0.2 days. The next
in importance is the R\o mer light-time effect (LITE) with the
amplitude of 0.022 days due to the change in the distance to the
stars "A" and "B" when they move in a common orbit. Then there are
slightly smaller nutation fluctuations with an amplitude of about
0.015 days. The problem was solved using the programs specially
developed for this case taking into account our experience in
searching for invisible satellites in eclipsing systems
\citet{2017CoSka..47...29V}, \citet{2012IAUS..282...89V},
\citet{2010ASPC..435..323V}, \citet{2011IBVS.5976....1V},
\citet{2015ASPC..496..109V}. The search for a solution was carried
out by the method of successive approximations. Both the
parameters of the apsidal rotation and the orbit of the third body
were searched simultaneously. In this case, both external orbits
for components "A" and "B" had to be identical and differ only in
the longitude of the periastron (by 180 degrees) and the
amplitudes of the LITE - due to the difference in the masses of
both systems. It was the observations of the "A" component that
made it possible to fix the parameters of the outer orbit most
reliably, since for it, the number of observed minima is maximal
and they are distributed over a significant time interval of 34
years. For the apsidal rotation of both eclipsing systems, the
parameters presented in Table~\ref{apsidal} were obtained.
\begin{figure}
\centering
\includegraphics[width=\hsize]{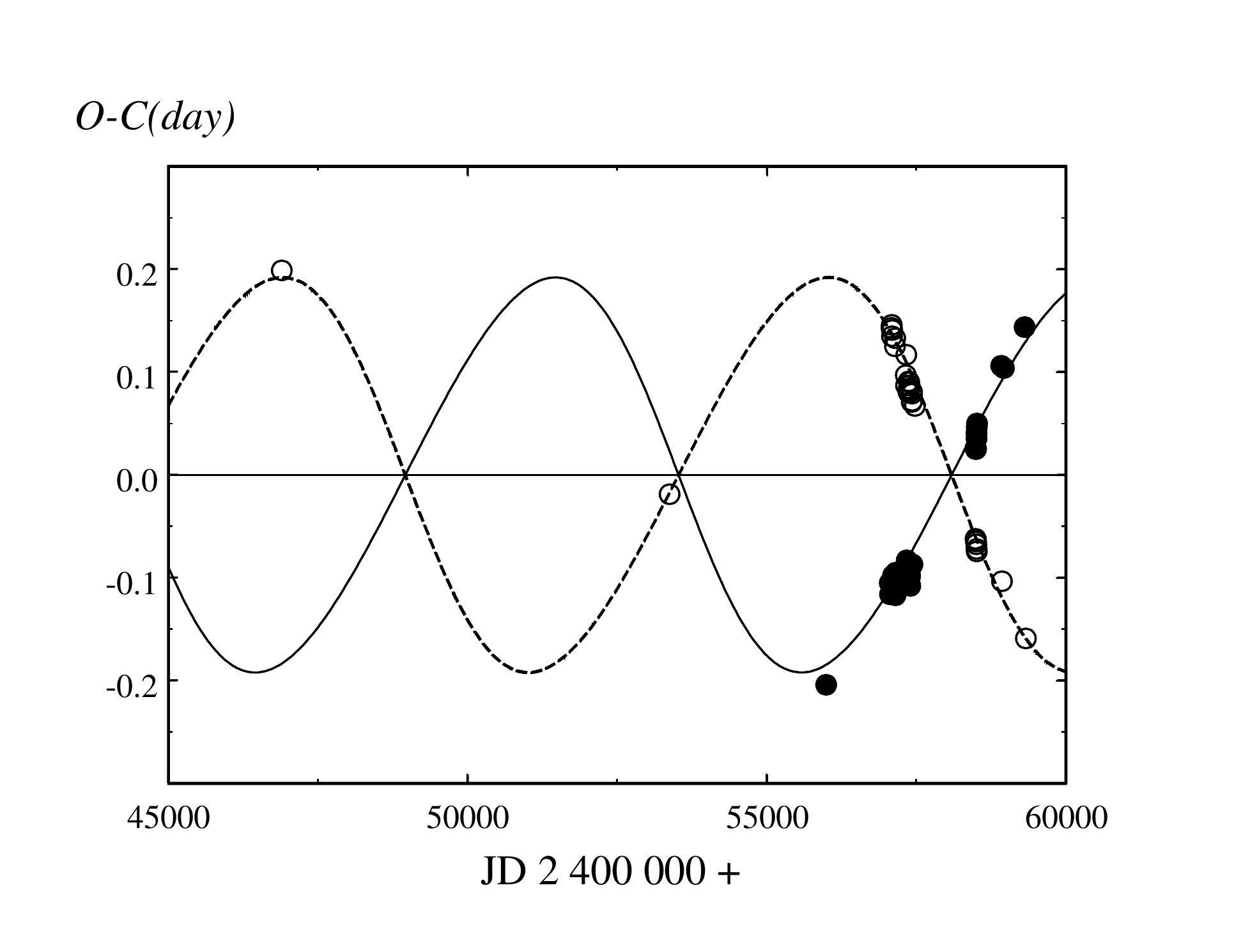}
\caption{The $O~-~C$ residuals calculated with the same period for
the primary and the secondary minima for "A" component. The
primary and the secondary minima are shown as open circles. The
solid and dashed lines correspond to detected apsidal motion for
the primary and the secondary minima, respectively.
              }
\label{Apsid_A}
\end{figure}
\begin{figure}
\centering
\includegraphics[width=\hsize]{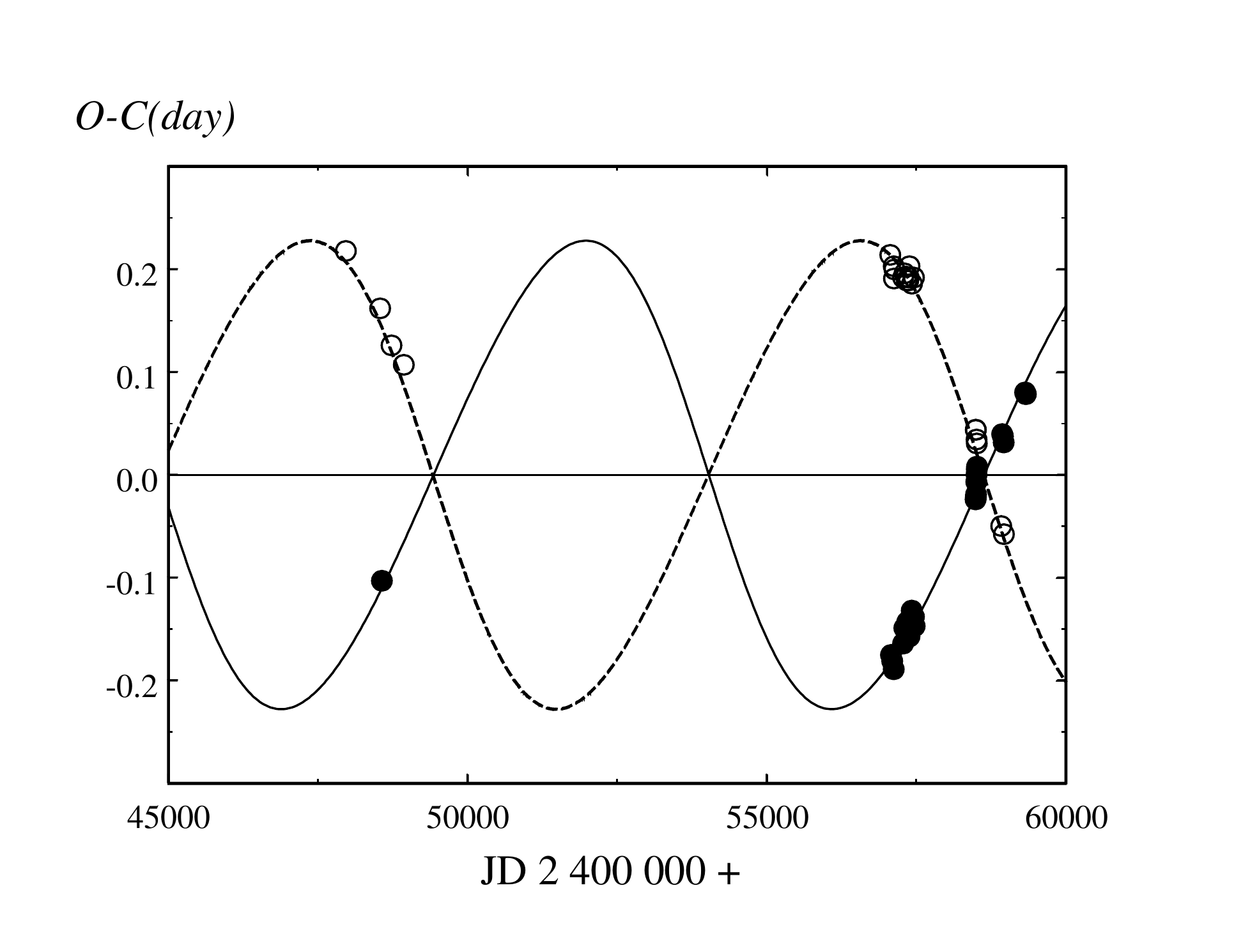}
\caption{The $O~-~C$ residuals calculated with the same period for
the primary and the secondary minima for "B" component. The
primary and the secondary minima are shown as open circles. The
solid and dashed lines correspond to detected apsidal motion for
the primary and the secondary minima, respectively.
              }
\label{Apsid_B}
\end{figure}
\begin{figure}
\centering
\includegraphics[width=\hsize]{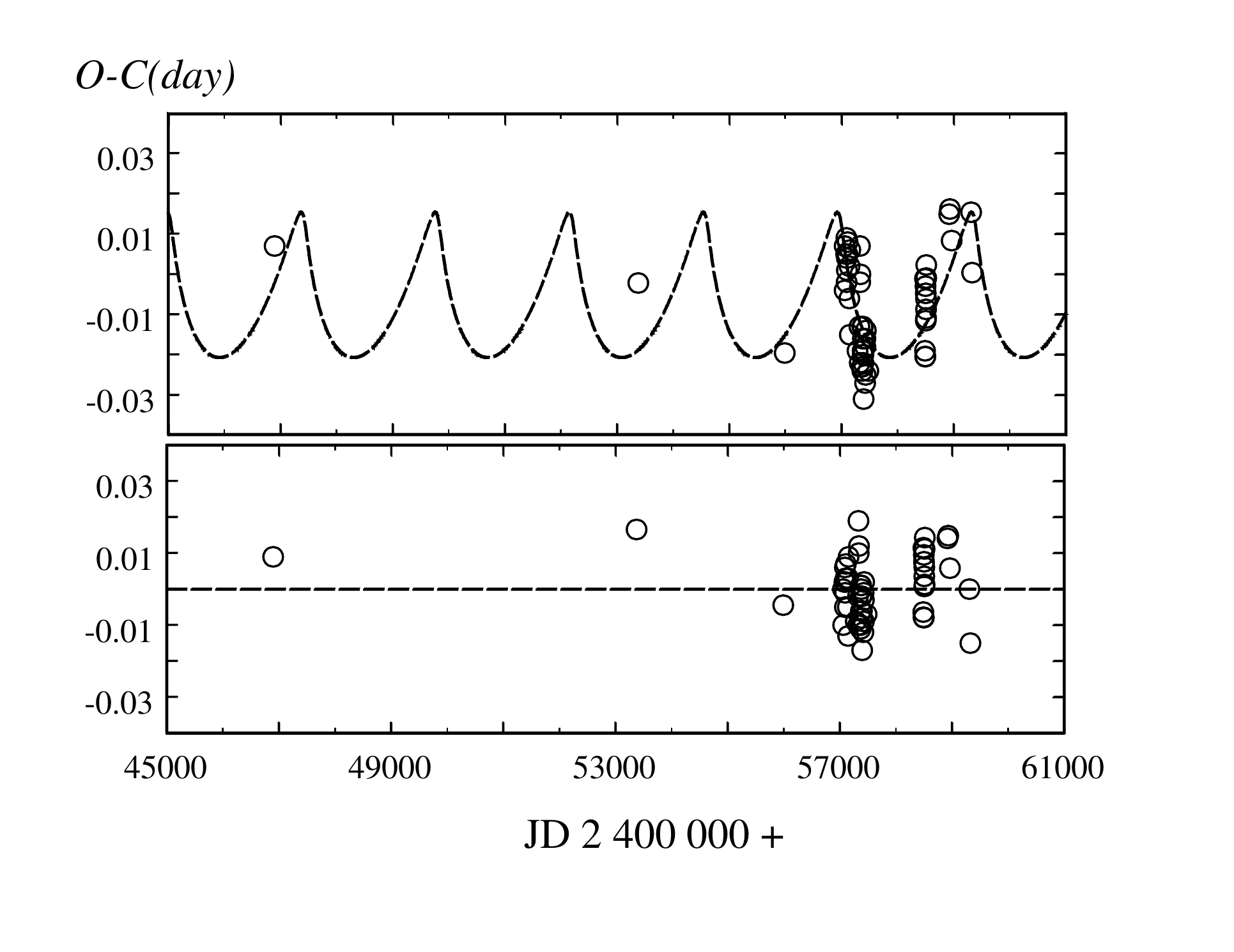}
\caption{Top panel: The $O~-~C$ graph after subtracting the
apsidal movement for the "A" component. The dashed line
corresponds to the parameters of the external orbit of the "A"
component around the center of gravity of "A"+"B" system. Lower
panel: residual deviations after subtracting the apsidal and
orbital movements. The residual scatter is associated with the
orbit nutation in the periastron longitude.
              }
\label{3body_A}
\end{figure}
\begin{figure}
\centering
\includegraphics[width=\hsize]{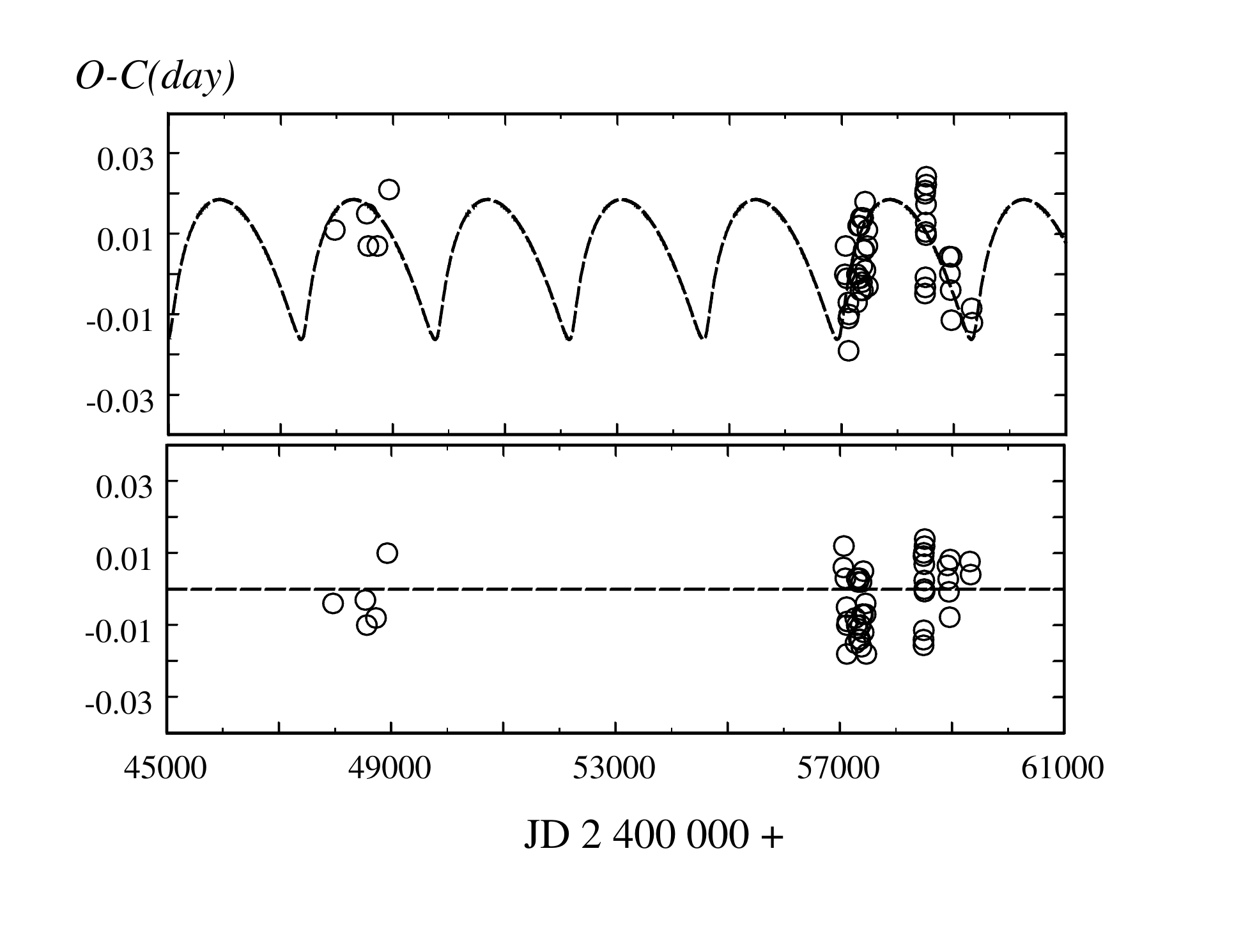}
\caption{Top panel: The $O~-~C$ graph after subtracting the
apsidal movement for the "B" component. The dashed line
corresponds to the parameters of the external orbit of the "B"
component around the center of gravity of "A"+"B" system. Lower
panel: residual deviations after subtracting the apsidal and
orbital movements. The residual scatter is associated with the
orbit nutation in the periastron longitude.
              }
\label{3body_B}
\end{figure}
After deduction of the apsidal rotation, the graphs of which are
shown in Figs.~\ref{Apsid_A} and \ref{Apsid_B}, the parameters of
the third body's orbit that most closely match the observed course
of the $O~-~C$ residuals (see Figs.~\ref{3body_A} and
\ref{3body_B}), are as follows:
    \begin{center}
    $P_3$~=~2390$^d\pm 40^d$ or 6.54 years \\
    $T_0$~=~J.D.~2,454,600 $\pm 20$ \\
    $(asin i)_A$~=~3.47$\pm 0.08$ AU \\
    $(asin i)_B$~=~3.34$\pm 0.08$ AU \\
    $e$ = 0.71 $\pm 0.03$, \\
    $\omega_A$ = $128^\circ\pm 3^\circ$, \\
    $\omega_B$ = $308^\circ\pm 3^\circ$, \\
    $f(M_3)_A$~=~0.9730$\pm 0.0005 M_\odot$ \\
    $f(M_3)_B$~=~0.8650$\pm 0.0005 M_\odot$ \\
    \end{center}
The errors are underestimated, they correspond to the specific
model, but there could exist several models which correspond to
their own local minima. So the parameters presented will not
necessarily be final. The period of the outer orbit is not fully
covered by observations yet. Our study of the entire area of
feasible solutions showed that there is a set of parameters which
differ from each other beyond the errors presented. The real
period of the outer orbit surely falls in 5.9-7.7 years interval.
We can also estimate that the ratio of the masses of the system
"A" to system "B" is approximately 1.0. This result is in good
agreement with photometric solutions, in which the luminosities of
components "A" and "B" are equally divided. Application of
Kepler's third law to the obtained parameters of external orbit
gives the total value of the mass of the entire system as
$M(sin~i)^3$~=~6.6~$M_\odot$. The solutions to the radial-velocity
curves assume $M$~=~13.1~$M_\odot$ (the inclination angles of the
orbits are known with high accuracy from the photometric
solutions). Hence, we conclude that $sin~i$~$\approx$~0.8, i.e.,
the angle of inclination of the outer orbit is close to 50 degrees
to the line of site. For the final conclusions about the external
orbit of the system further observations are required. We cannot
rule out that the above set of parameters of the outer orbit
corresponds to one of the local minima of the space of solutions.

\begin{figure}
\centering
\includegraphics[width=\hsize]{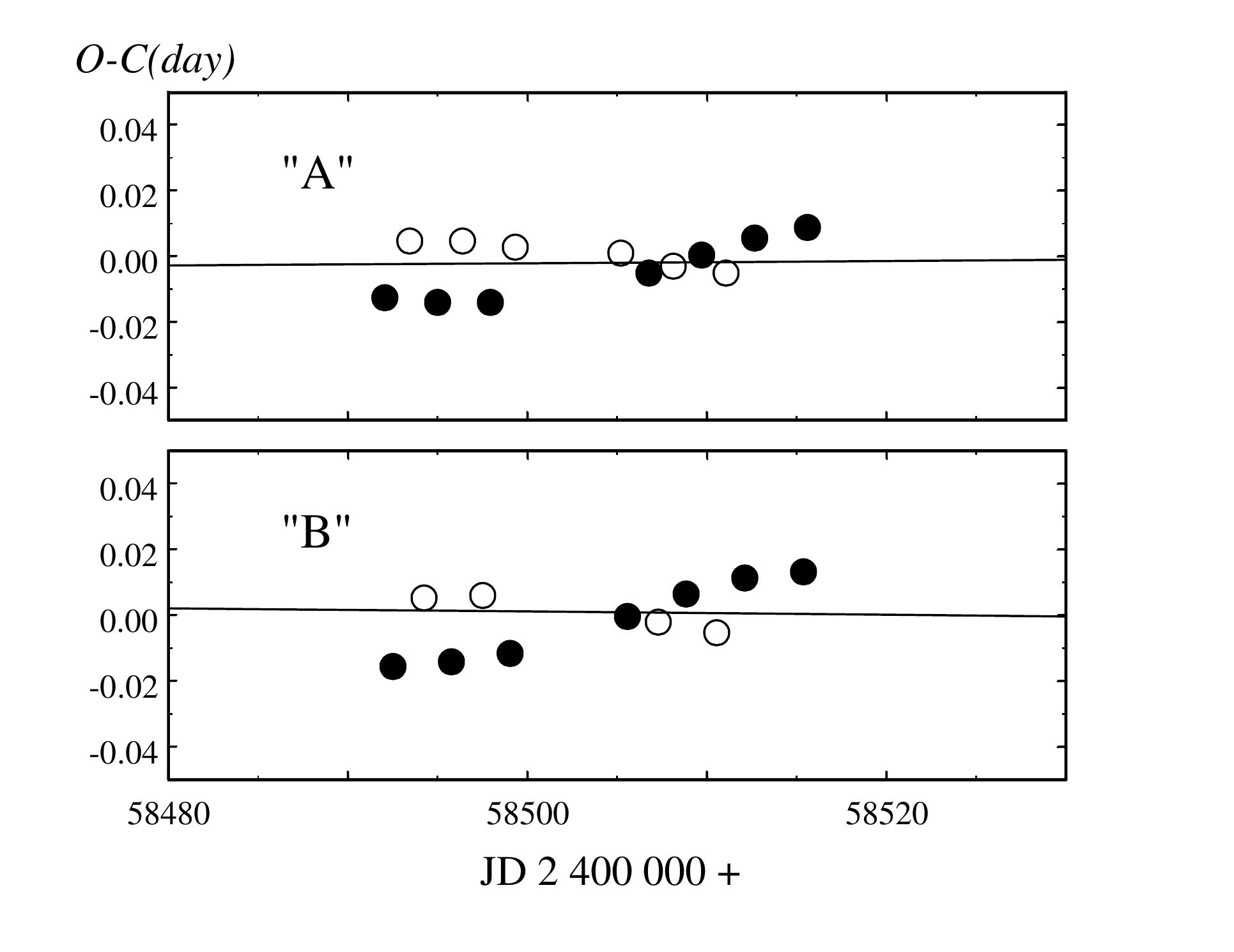}
\caption{Top panel: Figs.~\ref{3body_A}, \ref{3body_B} on an
enlarged scale for the TESS observations. Filled and open circles
present the primary and the secondary minima, respectively. The
solid oblique line corresponds to the R\o mer effect due to the
orbital motion of the "A" and "B" components. The orbit nutation
can explain observed oscillations in antiphase for the primary and
the secondary minima of the components "A" (upper panel) and "B"
(lower panel).
              }
\label{nutation}
\end{figure}

A large-scale $O-C$ plots for individual TESS timings is shown in
Fig.~\ref{nutation}. It can be seen that the primary and the
secondary minima for each of the "A" and "B" components slant in
antiphase. Such a picture can be obtained if we assume that the
elliptical orbits of each of the "A" and "B" components wobble
within one degree in the longitude of the periastron. We estimated
the periods of such wobbling at 60 days, and for both eclipsing
stars they are approximately the same. It turns out that in
addition to the usual effects observed in double elliptical
systems, such as apsidal motion and the LITE due to the influence
of the third body, nutation can also be present.

We do not give here the calculations of the theoretically expected
apsidal rotation. For such orbits and masses of the components, it
is obviously smaller than the measured values. The reasons for the
rapid apsidal rotation must be related to the mutual influence of
both systems on each other. Synchronization is clearly observed in
the system in the orbital, nutation and apsidal motions of the
components "A" and "B": the periods of the eclipsing pairs are
$P_A/P_B$=0.9, the periods of nutation are approximately equal,
the apsidal periods are also the same and $U/P_3 \approx 4$. The
orbital planes of both eclipsing systems are collinear. It can be
concluded that this multiple system has already reached an
equilibrium state and is stable.

\subsection{Conclusions}

We have revealed the nature of the unusual quadruple eclipsing
system BU~CMi, consisting of two eclipsing binaries. The periods
of both binaries are tied up by a resonance of 9/10. A fast
apsidal rotation is observed in each binary. We built a
preliminary orbit in which both binaries rotate around a common
center of mass. We found the physical characteristics of all four
stars: temperatures, sizes, masses. All stars appear to be very
young, aged up to 200 million years. A preliminary comparison of
the obtained parameters with theoretical models indicates that the
chemical composition of the stars does not correspond to the solar
one, and the secondary components are younger than the primary
ones. This fact is not surprising in the case of young stars.
Under the condition of common origin, less massive stars should
set on the ZAMS later than more massive ones. Further intensive
photometric and spectral observations are needed to refine the
parameters of the mutual orbit and to study nutation. We estimate
the angular distance between the "A" and "B" components at the
current moment to be about 0.01". So, it is possible to directly
measure the components of a binary star using speckle
interferometry. BU~CMi system is very interesting from the point
of view of stellar formation and evolution theory, as well as for
studying the dynamics of close multiple systems.

\begin{acknowledgements}

The observations were fulfilled with the 60-cm reflector of the
Simeiz observatory of INASAN. This work has used the SIMBAD
database of the Strasbourg Center for Astronomical Data (France)
and the bibliographic reference service of the ADS (NASA, USA).
This document includes data collected by the TESS mission. Funding
for the TESS mission is provided by NASA's Science Missions
Office. We are grateful to the staff members of the Astronomical
Institute of the Slovak Academy of Sciences T.~Pribulla and
R.~Kom\v{z}ik for the spectroscopic observations carried out at
our request.

This work was supported by a scholarship of Slovak Academic
Information Agency SAIA (ASK, IMV), RNF grant 14-12-00146 (IMV),
RFBR grant  11-02-01213a (IMV), and
the Slovak Research and Development Agency under the contract No.
APVV-15-045 (DC) and grant VEGA 2/0031/18 (DC).

\end{acknowledgements}
%
%---------------------------------------------------------
%
\newpage
%-------------Table1----------------
\begin{table}
\begin{center}
\caption{Observation log.} \label{journal}
\centering
\begin{tabular}{|c|c|c|c|}
\hline
Year     & Number of      & Photometric & Telescope \\
         & obs. nights    & passband    &           \\
\hline
2012     &  7             &    $V$          & 15-cm Maksutov \\
2020     & 20             &  $UBV$          & Zeiss-600 \\
2021     &  4             &    $V$          & Zeiss-600 \\
\hline
\end{tabular}
\end{center}
\end{table}
\newpage
\begin{table}
\caption{Stellar magnitudes of BU~CMi on plateau and standard star
in Johnson system.} \label{magnitudes}
\bigskip
\begin{tabular}{|c|c|c|c|}
\hline
Star           &  $V$    &  $U-B$  &  $B-V$  \\
\hline
BU~CMi         &  6.419  & -0.070  & -0.035  \\
               &  0.002  &  0.006  &  0.001  \\
HD64963        &  8.228  &  0.044  &  0.061  \\
               &  0.002  &  0.007  &  0.002  \\
[1mm] \hline
\end{tabular}
\end{table}
\begin{table}
\caption{ Absolute and relative parameters of BU~CMi.}
\label{masses}
\bigskip
\begin{tabular}{|c|c|c|}     % 3 columns
\hline
 Parameter     & Component "A"       &  Component "B"      \\
\hline
$r_1$          & $0.171(5)$    & $0.146(5)$    \\
$r_2$          & $0.123(5)$    & $0.129(5)$    \\
$i^\circ$      & $83.89(5)$     & $83.40(5)$    \\
$e$            & $0.204(7)$    & $0.218(7)$    \\
$\omega^\circ$ & $121.4(1)$      & $96.05(10)$       \\
$l_{1v}$       & $0.336(1)$    & $0.288(1)$    \\
$l_{2v}$       & $0.163(1)$    & $0.213(1)$    \\
$l_{1b}$       & $0.335(2)$   & $0.288(2)$    \\
$l_{2b}$       & $0.164(2)$   & $0.213(2)$    \\
$l_{1u}$       & $0.343(3)$   & $0.289(3)$    \\
$l_{2u}$       & $0.158(3)$   & $0.210(3)$    \\
$T_1$, K       & $10130(80)$     & $10180(80)$     \\
$T_2$, K       & $ 9740(80)$     & $9890(80)$     \\
B.C.$_1$      & $-0.277$            &  $-0.286$            \\
B.C.$_2$      & $-0.196$            &  $-0.227$            \\
\hline
$M_1$, M$_\odot$  & $3.40(10)$      & $3.29(10)$   \\
$M_2$, M$_\odot$  & $3.11(10)$      & $3.29(10)$   \\
$R_1$, R$_\odot$  & $2.51(5)$       & $2.31(5)$    \\
$R_2$, R$_\odot$  & $1.80(5)$       & $2.04(5)$    \\
lg $g_1$, cm/sec$^2$  & $4.088(10)$ & $4.148(10)$    \\
lg $g_2$, cm/sec$^2$  & $4.338(10)$ & $4.259(10)$     \\
$a$,  R$_\odot$   & $16.1(1)$       & $17.3(1)$       \\
$\pi$"$_{ph}$       & $0.00407(5)$  & $0.00407(5)$ \\
[1mm] \hline
$\omega^\circ$ refers to JD 2,458,921
\end{tabular}
\end{table}
\newpage
%

%
%---------------------Table 8-------------------------
\begin{table}
\caption{: Elements of the apsidal motion of BU~CMi components.
Values of eccentricity are fixed from the photometric solutions.}
\label{apsidal}
\bigskip
\begin{tabular}{|c|c|c|}
\hline
Parameter               &    "A"      &     "B"      \\
\hline
T0[HJD]                 & 2458092.265 & 2458636.080  \\
$P_s$ [days]            & 2.939550(5) & 3.262227(5)   \\
$e$                     & 0.20425     & 0.2181       \\
$\dot\omega$ [deg/year] & 14.4(1)     & 14.3(1)      \\
$\omega_0$[deg]         & 90          & 90           \\
$U$ [year]              & 25.0(2)     & 25.2(2)      \\
\hline
\end{tabular}
\end{table}
%
%
%---------------------Table 9-------------------------
\begin{table}
\caption{Radial velocities of BU~CMi components.}
\label{radial}
\bigskip
\begin{tabular}{|c|c|c|c|c|}
\hline
 HJD 2,400,000+     &  "A" pri.   &   "A" sec. &  "B" pri.   &   "B" sec. \\
                    &  km/sec         &   km/sec         &  km/sec         &   km/sec         \\
\hline
59163.6152   &  -30  &   40  &  110  &  -100  \\
59164.6629   &  125  & -120  &  -45  &   35   \\
59166.6279   &   25  &  -45  &   80  &   -50  \\
59177.6483   & -175  &  145  &   10  &    70  \\
59178.6138   &    5  &  -45  &  -85  &   115  \\
59179.6430   &   70  & -115  &  110  &   -45  \\
59180.5556   & -180  &  145  &  105  &   -75  \\
59185.6611   &   55  & -100  &   -5  &    55  \\
59196.5649   &   65  & -110  &  155  &   -70  \\
59197.5586   &   -5  &  -40  &  -80  &   165  \\
59203.5284   &  -25  &   25  &  130  &   -30  \\
59216.4063   &  -40  &   55  &  165  &   -55  \\
59224.4806   & -145  &  150  &  -25  &   100  \\
59226.4283   &  130  & -115  &   75  &   -20  \\
59246.3586   &   60  &  -95  &  -95  &   165  \\
59246.3844   &   70  &  -95  &  -95  &   170  \\
\hline
\end{tabular}
\end{table}
\begin{table}
\caption{The primary minima timings of  "A" component.}
\label{timings_A_pri}
\bigskip
\begin{tabular}{|l|l|l|l|c|}
\hline
HJD            & $(O - C)_1$, & $(O-C)_2$ & $(O - C)_3$    &  Note \\
(2,400,000+)   & days     &  days      &   days      &         \\
\hline
 55993.22212 & --0.20418 & --0.01954 & --0.00447 & 4 \\
 57057.439   & --0.105   &  +0.007   &  +0.000   & 5 \\
 57060.367   & --0.116   & --0.004   & --0.010   & 5 \\
 57098.591   & --0.106   &  +0.001   & --0.001   & 5 \\
 57107.414   & --0.102   &  +0.005   &  +0.003   & 5 \\
 57110.357   & --0.098   &  +0.008   &  +0.007   & 5 \\
 57151.492   & --0.117   & --0.015   & --0.013   & 5 \\
 57160.332   & --0.095   &  +0.006   &  +0.009   & 5 \\
 57286.721   & --0.108   & --0.019   & --0.009   & 5 \\
 57336.715   & --0.086   & --0.002   &  +0.010   & 5 \\
 57342.596   & --0.083   & --0.000   &  +0.012   & 5 \\
 57380.802   & --0.092   & --0.013   &  +0.001   & 5 \\
 57386.674   & --0.099   & --0.020   & --0.006   & 5 \\
 57389.618   & --0.094   & --0.016   & --0.002   & 5 \\
 57392.552   & --0.100   & --0.022   & --0.008   & 5 \\
 57395.494   & --0.098   & --0.020   & --0.006   & 5 \\
 57398.423   & --0.108   & --0.031   & --0.017   & 5 \\
 57436.658   & --0.087   & --0.014   &  +0.002   & 5 \\
 58492.06999 &  +0.02619 & --0.01901 & --0.00637 & 6 \\
 58495.00849 &  +0.02514 & --0.02038 & --0.00782 & 6 \\
 58497.94830 &  +0.02540 & --0.02045 & --0.00796 & 6 \\
 58506.77686 &  +0.03531 & --0.01152 &  +0.00076 & 6 \\
 58509.72214 &  +0.04104 & --0.00612 &  +0.00608 & 6 \\
 58512.66719 &  +0.04654 & --0.00094 &  +0.01119 & 6 \\
 58515.61030 &  +0.05010 &  +0.00229 &  +0.01435 & 6 \\
 58921.32414 &  +0.10604 &  +0.01491 &  +0.01409 & 7 \\
 58965.41533 &  +0.10398 &  +0.00838 &  +0.00585 & 7 \\
 59312.3220  &  +0.1438  &  +0.0154  & --0.0000  & 7 \\
\hline
\end{tabular}
\end{table}
\begin{table}
\caption{The secondary minima timings of  "A" component.}
\label{timings_A_sec}
\bigskip
\begin{tabular}{|l|l|l|l|c|}
\hline
HJD            & $(O - C)_1$, & $(O-C)_2$ & $(O - C)_3$    &  Note \\
(2,400,000+)   & days     &  days      &   days      &         \\
\hline
 46897.191   &  +0.199   &  +0.007   &  +0.009   & 1 \\
 53378.6805  & --0.0188  & --0.0022  &  +0.0166  & 3 \\
 57085.615   &  +0.143   &  +0.005   &  +0.002   & 5 \\
 57091.497   &  +0.146   &  +0.009   &  +0.006   & 5 \\
 57094.425   &  +0.135   & --0.002   & --0.005   & 5 \\
 57097.371   &  +0.141   &  +0.004   &  +0.002   & 5 \\
 57144.388   &  +0.125   & --0.006   & --0.005   & 5 \\
 57147.335   &  +0.133   &  +0.002   &  +0.003   & 5 \\
 57323.673   &  +0.097   & --0.013   & --0.002   & 5 \\
 57329.542   &  +0.087   & --0.022   & --0.010   & 5 \\
 57332.511   &  +0.117   &  +0.007   &  +0.019   & 5 \\
 57367.750   &  +0.082   & --0.023   & --0.010   & 5 \\
 57370.690   &  +0.082   & --0.023   & --0.010   & 5 \\
 57373.628   &  +0.080   & --0.024   & --0.011   & 5 \\
 57376.578   &  +0.091   & --0.013   &  +0.001   & 5 \\
 57379.511   &  +0.084   & --0.019   & --0.006   & 5 \\
 57382.456   &  +0.090   & --0.013   &  +0.000   & 5 \\
 57385.393   &  +0.087   & --0.016   & --0.002   & 5 \\
 57423.591   &  +0.071   & --0.027   & --0.012   & 5 \\
 57426.541   &  +0.081   & --0.016   & --0.001   & 5 \\
 57429.478   &  +0.079   & --0.018   & --0.003   & 5 \\
 57432.411   &  +0.072   & --0.025   & --0.009   & 5 \\
 57479.438   &  +0.067   & --0.024   & --0.007   & 5 \\
 58493.45422 & --0.06213 & --0.00109 &  +0.01151 & 6 \\
 58496.39330 & --0.06260 & --0.00113 &  +0.01140 & 6 \\
 58499.33059 & --0.06486 & --0.00296 &  +0.00950 & 6 \\
 58505.20707 & --0.06748 & --0.00473 &  +0.00758 & 6 \\
 58508.14223 & --0.07187 & --0.00869 &  +0.00355 & 6 \\
 58511.07922 & --0.07443 & --0.01083 &  +0.00134 & 6 \\
 58934.34536 & --0.10349 &  +0.01618 &  +0.01486 & 7 \\
 59331.1291  & --0.1590  &  +0.0004  & --0.0150  & 7 \\
\hline
\end{tabular}
\end{table}
\begin{table}
\caption{The primary minima timings of "B" component.}
\label{timings_B_pri}
\bigskip
\begin{tabular}{|l|l|l|l|c|}
\hline
HJD            & $(O - C)_1$, & $(O - C)_2$ & $(O - C)_3$    &  Note \\
(2,400,000+)   & days     &  days      &   days      &         \\
\hline
 48568.744   & --0.103   &   +0.007   & --0.010   & 2 \\
 57076.560   & --0.175   &   +0.007   &  +0.012   & 5 \\
 57099.390   & --0.181   &  --0.001   &  +0.003   & 5 \\
 57125.480   & --0.189   &  --0.011   & --0.010   & 5 \\
 57278.830   & --0.164   &   +0.000   & --0.008   & 5 \\
 57301.680   & --0.149   &   +0.012   &  +0.003   & 5 \\
 57327.780   & --0.147   &   +0.012   &  +0.002   & 5 \\
 57350.620   & --0.143   &   +0.014   &  +0.003   & 5 \\
 57386.490   & --0.157   &  --0.004   & --0.016   & 5 \\
 57399.550   & --0.146   &   +0.006   & --0.007   & 5 \\
 57412.600   & --0.145   &   +0.006   & --0.007   & 5 \\
 57422.400   & --0.132   &   +0.018   &  +0.005   & 5 \\
 57461.540   & --0.138   &   +0.007   & --0.007   & 5 \\
 57474.580   & --0.147   &  --0.003   & --0.018   & 5 \\
 58492.51817 & --0.02384 &  --0.00479 & --0.01560 & 6 \\
 58495.78232 & --0.02192 &  --0.00330 & --0.01403 & 6 \\
 58499.04752 & --0.01895 &  --0.00076 & --0.01142 & 6 \\
 58505.58412 & --0.00680 &   +0.01052 &  +0.00002 & 6 \\
 58508.85361 &  +0.00046 &   +0.01735 &  +0.00693 & 6 \\
 58512.12118 &  +0.00581 &   +0.02226 &  +0.01191 & 6 \\
 58515.38576 &  +0.00816 &   +0.02418 &  +0.01391 & 6 \\
 58936.24472 &  +0.03984 &   +0.00015 &  +0.00282 & 7 \\
 58949.29126 &  +0.03747 &  --0.00393 &  -0.00077 & 7 \\
 58962.33437 &  +0.03167 &  --0.01144 & --0.00779 & 7 \\
 59321.22804 &  +0.08037 &  --0.00850 &  +0.00765 & 7 \\
 59334.27505 &  +0.07847 &  --0.01201 &  +0.00404 & 7 \\
\hline
\end{tabular}
\end{table}
\begin{table}
\caption{The secondary minima timings of "B" component.}
\label{timings_B_sec}
\bigskip
\begin{tabular}{|l|l|l|l|c|}
\hline
HJD            & $(O - C)_1$, & $(O-C)_2$ & $(O - C)_3$    &  Note \\
(2,400,000+)   & days     &  days      &   days      &         \\
\hline
 47967.223   &  +0.218   &  +0.011   &  -0.004   & 2 \\
 48541.319   &  +0.162   &  +0.015   &  -0.003   & 2 \\
 48730.493   &  +0.126   &  +0.007   & --0.008   & 2 \\
 48935.994   &  +0.107   &  +0.021   &  +0.010   & 2 \\
 57065.570   &  +0.214   & --0.000   &  +0.006   & 5 \\
 57124.280   &  +0.203   & --0.007   & --0.005   & 5 \\
 57127.530   &  +0.191   & --0.019   & --0.018   & 5 \\
 57137.325   &  +0.200   & --0.010   & --0.009   & 5 \\
 57280.855   &  +0.192   & --0.007   & --0.015   & 5 \\
 57303.695   &  +0.196   & --0.001   & --0.010   & 5 \\
 57329.790   &  +0.193   & --0.001   & --0.011   & 5 \\
 57342.835   &  +0.189   & --0.004   & --0.014   & 5 \\
 57365.670   &  +0.189   & --0.002   & --0.014   & 5 \\
 57375.460   &  +0.192   &  +0.002   & --0.010   & 5 \\
 57388.520   &  +0.203   &  +0.014   &  +0.002   & 5 \\
 57427.650   &  +0.186   &  +0.001   & --0.012   & 5 \\
 57463.540   &  +0.192   &  +0.011   &  -0.004   & 5 \\
 58494.25583 &  +0.04391 &  +0.01997 &  +0.00920 & 6 \\
 58497.51828 &  +0.04413 &  +0.02079 &  +0.01010 & 6 \\
 58507.29522 &  +0.03439 &  +0.01284 &  +0.00238 & 6 \\
 58510.55370 &  +0.03064 &  +0.00969 &  -0.00069 & 6 \\
 58921.5141  & --0.0496  &  +0.0044  &  +0.0066  & 7 \\
 58967.1771  & --0.0577  &  +0.0043  &  +0.0082  & 7 \\
\hline
\end{tabular}
\end{table}
FOOTNOTES TO TABLES~\ref{timings_A_pri}, \ref{timings_A_sec}, \ref{timings_B_pri}, \ref{timings_B_sec}: \\
The errors of minima times approximately correspond to 5 units of the last significant digit.\\
$(O - C)_1$ -- deviation from the linear formula.\\
$(O - C)_2$ -- residuals after the deduction of the apsidal movement.\\
$(O - C)_3$ -- residuals after the deduction of the apsidal movement and the light-time effect.\\
Notes: \\
(1) -- data obtained from the SAI catalogue,\\
(2) -- Hipparcos observations, \\
(3) -- \citet{2008IBVS.5843....1O},\\
(4) -- CCD observations, Star\'a Lesn\'a, Slovakia,\\
(5) -- MASCARA observations,\\
(6) -- TESS observations,\\
(7) -- Observations with a photomultiplier, INASAN Simeiz
Observatory.\\
%
% for the bibliography, at the end

%
%------------------------------------------------------------------
\end{document}